\documentclass[conference]{IEEEtran}
\usepackage{alltt                                    
          , multirow
          , booktabs
          , listings
          , graphicx
          ,float
	,cite
          ,verbatim
         ,mathtools
	,url
	,amsmath
}
\usepackage[table]{xcolor}
\usepackage[numbers]{natbib}     
\usepackage{syntax}
\usepackage[ruled]{algorithm}
\usepackage{algpseudocode}
\usepackage{enumitem}
\usepackage{threeparttable}
\usepackage{framed}
\usepackage{hhline}
\usepackage{dingbat}
\usepackage{balance}

\usepackage{tikz}
\usepackage{verbatim}
\usetikzlibrary{arrows,shapes,backgrounds}

\tikzstyle{vertex}=[ellipse,fill=black!25,minimum size=20pt, inner sep=0pt]
\tikzstyle{edge} = [draw,thin,-]
\tikzstyle{glabel} = [text width=1cm,text centered,font=\bf]
\pgfdeclarelayer{bg}    
\pgfsetlayers{bg,main}  

\usepackage{expl3}
\ExplSyntaxOn
\newcommand\latinabbrev[1]{
  \peek_meaning:NTF . {
    #1\@}%
  { \peek_catcode:NTF a {
      #1., \@ }%
    {#1., \@}}}
\ExplSyntaxOff

\newcommand{\CASE}[1]{\STATE \textbf{case} #1\textbf{:} \begin{ALC@g}}
\newcommand{\ENDCASE}{\end{ALC@g}}

\newcommand{\DEFAULT}{\STATE \textbf{default:} \begin{ALC@g}}
\newcommand{\ENDDEFAULT}{\end{ALC@g}}
\newcommand{\DEFAULTLINE}[1]{\STATE \textbf{default:} }
\let\footnotesize\scriptsize

\algnewcommand{\LineComment}[1]{\State \(\triangleright\) #1}
\algdef{SE}[DOWHILE]{Do}{doWhile}{\algorithmicdo}[1]{\algorithmicwhile\ #1}%

\newsavebox{\supbox}
\newcommand{\bsup}{\begin{lrbox}{\supbox}$\tt\scriptstyle}
\newcommand{\esup}{$\end{lrbox}{}^{\usebox{\supbox}}}
\def\eg{\latinabbrev{e.g}}
\def\ie{\latinabbrev{i.e}}

\definecolor{ashgrey}{rgb}{0.7, 0.75, 0.71}
\definecolor{babyblue}{rgb}{0.54, 0.81, 0.94}

\definecolor{lightpurple}{rgb}{0.8,0.8,1}
\definecolor{codebg}{RGB}{255,255,255}
\definecolor{commentcolor}{RGB}{11,140,11}
\begin{document}
%

\title{STRICT: Information Retrieval Based Search Term Identification for Concept Location\vspace{-.3cm}}
%
%
%
%
%

\author{\IEEEauthorblockN{Mohammad Masudur Rahman  ~~~ Chanchal K. Roy}
\IEEEauthorblockA{Department of Computer Science, University of Saskatchewan, Canada\\
\{masud.rahman, chanchal.roy\}@usask.ca}
}

\maketitle

\begin{abstract}
During maintenance, software developers deal with numerous  change requests that are written in an unstructured fashion using natural language.
Such natural language texts illustrate the change requirement involving various domain related concepts. 
Software developers need to find appropriate search terms from those concepts so that they could locate the possible locations in the source code using a search technique. 
Once such locations are identified, they can implement the requested changes there.
Studies suggest that developers often perform poorly in coming up with good search terms for a change task. 
In this paper, we propose a novel technique--STRICT--that automatically identifies suitable search terms for a software change task by analyzing its task description using two information retrieval (IR) techniques-- TextRank and POSRank.
These IR techniques determine a term's importance based on not only its co-occurrences with other important terms but also its syntactic relationships with them.
Experiments using 1,939 change requests from eight subject systems report that STRICT can identify better quality search terms than baseline terms from 52\%--62\% of the requests with 30\%--57\% Top-10 retrieval accuracy which are promising.
Comparison with two state-of-the-art techniques not only validates our empirical findings and but also demonstrates the superiority of our technique.

\end{abstract}


\IEEEpeerreviewmaketitle

\section{Introduction}
During maintenance, software developers deal with numerous change requests as a part of 
software change implementation.
Identification of exact source location (\ie\ also called \emph{concept location}) during software code change (\eg\ new feature addition, bug fixation) is a major challenge, even for a medium sized system \cite{infer}. 
Change requests are often made by the users of a software system, and these requests are generally written using unstructured natural language texts \cite{kevicdict}. 
While the software users might be familiar with the application domain of a software system, they generally lack the idea of how a particular software feature is implemented in the source code.
Hence, a change request from them generally involves one or more \emph{``high level"} concepts (\ie\ functional requirement) related to the application domain. A developer needs to map these concepts to the relevant source locations within the project
for implementing the change \cite{kevicdict, textret}. 
Such mapping is possibly trivial for the developer who has substantial knowledge on the target project. 
However, developers involved in the maintenance might not be aware of the low-level architecture of the software project, and the design documents might not be available either \cite{ossdoc,measure}.
Thus, they often experience difficulties in identifying the exact source locations (\eg\ methods) to be changed. 
The mapping task generally starts with a search within the project which requires one or more suitable search terms. 
Previous studies \cite{kevic,vocaprob,sitir} suggest that on average, developers perform poorly in coming up with good search terms for a change task regardless of their experience. 
For example, \citet{kevic} report from a user study that only 12.20\% of the search terms from the developers were able to retrieve relevant results.
According to \citet{vocaprob}, there is a small chance (10\%--15\%) that developers guess the exact words used in the source code. 
One way to help them overcome this challenge is to automatically suggest suitable search terms for the change task at hand, and our work addresses this research problem.

Existing studies from relevant literature apply various lightweight heuristics \cite{kevic} and query reformulation strategies \cite{gayg, refoqus, shepherd,infer}. They also perform different query quality analyses \cite{qquality,qeffect,specificity,refoqus} and data mining activities \cite{kevicdict,ccmapping,infer}. 
However, most of these approaches expect a developer to provide the initial search query which they can improve upon. 
Unfortunately, preparing such a query is often a non-trivial task for the developers as well \cite{kevic,vocaprob}.
One can think of the whole change request as the initial query. However, this might lead the developers to major query reformulation tasks \cite{gayg} or suboptimal solutions.
\citeauthor{kevic} propose the only heuristic model for automatically identifying initial search terms for a change task where they consider different heuristics associated with \emph{frequency, location, parts of speech} and \emph{notation} of the terms from the change request. Although their model is found  promising 
according to the preliminary evaluation, it suffers from two major limitations. 
First, their model is neither trained using a large dataset (\ie\ uses only 20 change requests) nor cross-validated using the change requests from multiple software projects (\ie\ uses only one project). 
Thus, the model is yet to be matured and reliable.
Second, \emph{tf--idf} is the most dominating feature of their model which is subject to the size of the change request dataset for \emph{inverse document frequency (idf)} calculation.
Hence, their model is likely to be affected significantly by the size of the dataset. That is, it might require frequent re-training to keep itself useful for search term identification. 
Thus, an approach that overcomes such limitations and yet can identify appropriate search terms from a given change request is warranted.

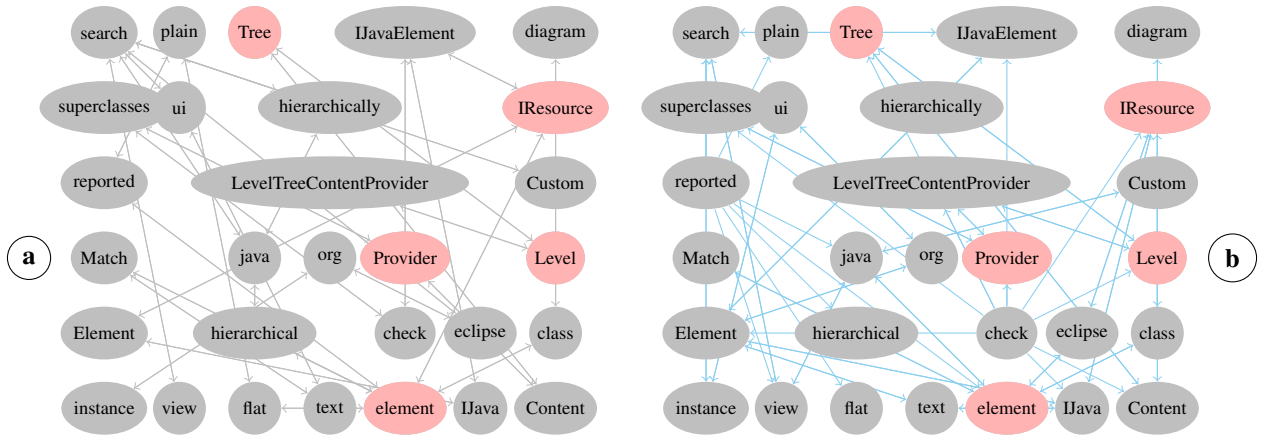
\begin{figure*}
\centering
\begin{tikzpicture}[scale=1, auto,swap]
    \foreach \pos/\name in {
{(1,1)/instance},
{(1,2)/Element},
{(1,3)/Match},
{(1,5)/superclasses},
{(1,6)/search},
{(2,1)/view},
{(3,3)/java},
{(2,5)/ui},
{(2,6)/plain},
{(3,1)/flat},
{(3,2)/hierarchical},
{(1,4)/reported},
{(3,6)/Tree},
{(4,1)/text},
{(7,2)/class},
{(5,1)/element},
{(5,6)/IJavaElement},
{(4,3)/org},
{(7,4)/Custom},
{(4,5)/hierarchically},
{(5,2)/check},
{(6,1)/IJava},
{(6,2)/eclipse},
{(4,4)/LevelTreeContentProvider},
{(5,3)/Provider},
{(7,6)/diagram},
{(7,1)/Content},
{(7,3)/Level},
{(7,5)/IResource}}
 \node[vertex] (\name) at \pos {\scriptsize\name};

\node[vertex] at (5,1) [fill=red!30] {\scriptsize element};
\node[vertex] at (7,5) [fill=red!30] {\scriptsize IResource};
\node[vertex] at (5,3) [fill=red!30] {\scriptsize Provider};
\node[vertex] at (7,3) [fill=red!30] {\scriptsize Level};
\node[vertex] at (3,6) [fill=red!30] {\scriptsize Tree};

\begin{pgfonlayer}{bg}    
\foreach \source/ \dest  in {
search/Custom,
Custom/search,
search/hierarchically,
hierarchically/search,
hierarchically/java,
java/hierarchically,
java/search,
search/java,
search/view,
view/search,
instance/org,
org/instance,
org/eclipse,
eclipse/org,
eclipse/search,
search/eclipse,
search/ui,
ui/search,
ui/text,
text/ui,
text/Match,
Match/text,
Match/element,
element/Match,
element/IResource,
IResource/element,
IResource/IJavaElement,
IJavaElement/IResource,
IJavaElement/IJava,
IJava/IJavaElement,
IJava/Element,
Element/IJava,
element/class,
class/element,
class/diagram,
diagram/class,
element/reported,
reported/element,
reported/plain,
plain/reported,
plain/flat,
flat/plain,
flat/element,
element/flat,
element/hierarchical,
hierarchical/element,
hierarchical/java,
java/hierarchical,
LevelTreeContentProvider/Level,
Level/LevelTreeContentProvider,
Level/Tree,
Tree/Level,
Tree/Content,
Content/Tree,
Content/Provider,
Provider/Content,
Provider/superclasses,
superclasses/Provider,
superclasses/check,
check/superclasses,
check/IJavaElement,
IJavaElement/check,
Element/IResource,
IResource/Element}
\path[edge, ->, draw=lightgray] (\source) -- (\dest);
\end{pgfonlayer}

\node at (0,3) [circle,draw] (a) {\bf{a}};

\foreach \pos/\name in {
{(9,1)/instance},
{(9,2)/Element},
{(9,3)/Match},
{(9,5)/superclasses},
{(9,6)/search},
{(10,1)/view},
{(11,3)/java},
{(10,5)/ui},
{(10,6)/plain},
{(11,1)/flat},
{(11,2)/hierarchical},
{(9,4)/reported},
{(11,6)/Tree},
{(12,1)/text},
{(15,2)/class},
{(13,1)/element},
{(13,6)/IJavaElement},
{(12,3)/org},
{(15,4)/Custom},
{(12,5)/hierarchically},
{(13,2)/check},
{(14,1)/IJava},
{(14,2)/eclipse},
{(12,4)/LevelTreeContentProvider},
{(13,3)/Provider},
{(15,6)/diagram},
{(15,1)/Content},
{(15,3)/Level},
{(15,5)/IResource}}
 \node[vertex] (\name) at \pos {\scriptsize\name};

\node[vertex] at (13,1) [fill=red!30] {\scriptsize element};
\node[vertex] at (15,5) [fill=red!30] {\scriptsize IResource};
\node[vertex] at (13,3) [fill=red!30] {\scriptsize Provider};
\node[vertex] at (15,3) [fill=red!30] {\scriptsize Level};
\node[vertex] at (11,6) [fill=red!30] {\scriptsize Tree};

\begin{pgfonlayer}{bg}    
\foreach \source/ \dest  in {
search/view,
view/search,
view/java,
java/view,
java/Custom,
Custom/java,
IJavaElement/search,
search/IJavaElement,
search/instance,
instance/search,
instance/ui,
ui/instance,
ui/org,
org/ui,
org/Element,
Element/org,
Element/text,
text/Element,
text/IJava,
IJava/text,
IJava/IResource,
IResource/IJava,
IResource/eclipse,
eclipse/IResource,
eclipse/element,
element/eclipse,
element/Match,
Match/element,
diagram/class,
class/diagram,
class/element,
element/class,
java/element,
element/java,
reported/search,
reported/view,
reported/java,
reported/element,
reported/plain,
reported/flat,
reported/hierarchical,
IJavaElement/Element,
Element/IJavaElement,
Element/IJava,
IJava/Element,
IResource/Content,
Content/IResource,
Content/Tree,
Tree/Content,
Tree/Level,
Level/Tree,
Level/LevelTreeContentProvider,
LevelTreeContentProvider/Level,
LevelTreeContentProvider/Provider,
Provider/LevelTreeContentProvider,
Provider/superclasses,
superclasses/Provider,
check/IJavaElement,
check/Element,
check/IJava,
check/IResource,
check/Content,
check/Tree,
check/Level,
check/LevelTreeContentProvider,
check/Provider,
check/superclasses}
\path[edge, ->, draw=babyblue] (\source) -- (\dest);
\end{pgfonlayer}

\node at (16,3) [circle,draw] (b) {\bf{b}};

\end{tikzpicture}
\caption{Text Graphs of change request in Table \ref{table:ctask} -- (a) using word co-occurrences, and (b) using syntactic dependencies}
\label{fig:tgraph}
\vspace{-.5cm}
\end{figure*}

In this paper, we propose a novel query recommendation technique--STRICT--that automatically identifies and recommends good quality search terms from a change request for concept location.
We determine importance of each term by employing two graph-based term weighting algorithms (from information retrieval domain)--\emph{TextRank} and \emph{POSRank}--on the content of a change request, and then suggest the most important terms as the search query.
Both TextRank and POSRank are adaptations of \emph{PageRank} algorithm \cite{pagerank} for natural language texts, and they extract the most important terms from a change request by analyzing co-occurrences \cite{rada} and syntactic dependencies \cite{rada, blanco} among the terms respectively. 
Unlike \citet{kevic}, they are not subject to the size of subject systems under study, and also do not require any training.
In fact, our technique is highly suited for the target research problem from several perspectives.
First, our technique considers the content of a change request as a text graph (\eg\ Fig. \ref{fig:tgraph}) based on either term co-occurrences or syntactic dependencies rather than plain text.
Thus, it has a higher potential for revealing important semantic relationships among the terms which might lead one to choosing appropriate search terms.
Second, both TextRank and POSRank determine importance of a term in the graph (\ie\ change request) not only based on its connectivity but also by considering the weights (\ie\ importance) of the connected terms. 
That is, a term would be considered important only if it connected to other important terms \cite{rada,sameer}.
Thus, our technique has a better chance of locating the chunk of important terms from the change request which can be suggested as the search terms.

Table \ref{table:ctask} shows an example change request from \texttt{eclipse.jdt.ui} system that reports a concern with custom search result display in Eclipse IDE.
Our technique--STRICT--first converts the textual content of the request into two text graphs by capturing (a) co-occurrences of the terms in the request (\ie\ Fig. \ref{fig:tgraph}-(a)) and (b) dependencies among the terms due to their parts of speech (POS) (\ie\ Fig. \ref{fig:tgraph}-(b)) respectively. 
Then, it identifies the most important terms by recursively analyzing the topological characteristics (\ie\ term connectivity) of both graphs.
STRICT returns the following Top-5 search terms (\ie\ highlighted, Fig. \ref{fig:tgraph})-- \emph{`element', `IResource', `Provider', `Level'} and \emph{`Tree'}--which return the first correct result at the Top-1 position.
On the contrary, the baseline queries-- \emph{Title}, \emph{Description} and \emph{Title}+\emph{Description}-- return such result at 559$^{th}$, 71$^{st}$ and 211$^{th}$ positions. 
Thus, our suggested search terms (1) can provide a meaningful starting point for code search during code change, and (2) can potentially reduce the query reformulation efforts spent by the developers.  
It should be noted that this paper is a significantly extended version of our preliminary work on search term identification \cite{saner2015masud}. 
While the earlier work explores the potential of \emph{TextRank} using a limited dataset, this work (1) extends that idea by applying another appropriate term-weighting technique--\emph{POSRank}, (2) proposes a novel search term ranking algorithm, 
and (3) then evaluates and validates the technique extensively using a larger dataset \cite{strict}.

\begin{table}[!t]
\centering
\caption{An Example Change Request (Issue \#:303705, eclipse.jdt.ui)}\label{table:ctask}
\vspace{-.2cm}
\resizebox{3.5in}{!}{%
\begin{threeparttable}
\begin{tabular}{l| p{6cm}}
\hline
\textbf{Field} & \textbf{Content}\\
\hline
\hline
Title & [search] Custom search results not shown hierarchically in the java search results view \\
\hline
Descripton & Consider an instance of org.eclipse.search.ui.text.Match with an element that is neither an IResource nor an IJavaElement. It might be an element in a class diagram, for example.
When such an element is reported, it will be shown as a plain, flat element in the otherwise hierarchical java search results view. This is because the LevelTreeContentProvider and its superclasses only check for IJavaElement and IResource.
\\
\hline
\end{tabular}
\centering
\end{threeparttable}
}
\vspace{-.6cm}
\end{table}

Experiments using 1,939 change requests from eight subject systems (\ie\ three \emph{Apache} systems and five \emph{Eclipse} systems) report that our technique--STRICT--can
provide better quality search terms than 52\%--62\% of the baseline queries which is highly promising according to relevant literature \cite{refoqus,trconfig}.
Our suggested queries can retrieve relevant source code files for 30\%--57\% of the change tasks with about 30\% mean average precision@10 where the first relevant file is mostly found within the Top-4 positions, which are also promising \cite{antoniol,bavota}.   
Comparison with two state-of-the-art techniques--\citet{kevic} and \citet{rocchio}-- not only validates our empirical findings but also demonstrates the superiority of our technique. 
We thus make the following contributions:
\begin{itemize}[noitemsep]
\item A novel and promising search term identification technique--STRICT--that identifies good quality search terms for a change task from its change request.
\item Comprehensive evaluation of the technique using 1,939 change requests from eight subject systems and four state-of-the-art performance metrics.
\item Comprehensive validation of the technique using comparisons with two state-of-the-art techniques.
\end{itemize}



\section{Graph Based Term-Weighting}\label{sec:bg}

\begin{figure*}
\centering
\begin{tikzpicture}[scale=.9, auto,swap]

\node at (1,2.8) [circle,draw] (1) {\scriptsize{1}};
\node at (3,2.8) [circle,draw] (2) {\scriptsize{2}};
\node at (5,3.8) [circle,draw] (3) {\scriptsize{3}};
\node at (5.5,1.5) [circle,draw] (4) {\scriptsize{4}};
\node at (7,3.8) [circle,draw] (5) {\scriptsize{5}};
\node at (7.7,1) [circle,draw] (6) {\scriptsize{6}};
\node at (9,2.8) [circle,draw] (7) {\scriptsize{7}};
\node at (11,2.8) [circle,draw] (8) {\scriptsize{8}};

\begin{pgfonlayer}{bg}
\node[inner sep=0pt] (cr) at (0.1,2)
    {\includegraphics[width=.25in]{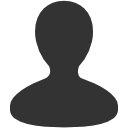}};
\node[inner sep=0pt] (cr) at (1,2)
    {\includegraphics[width=.5in]{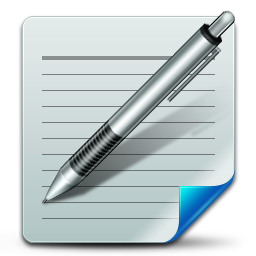}};
\node at (1,1) (b) {\scriptsize Software change};
\node at (1,0.7) (b) {\scriptsize request (\bf{input})};

\node[inner sep=0pt] (prep) at (3,2)
    {\includegraphics[width=.5in]{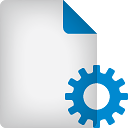}};
\node at (3,1) (b) {\scriptsize Preprocessing};
\draw[->,thick] (cr) -- (prep);
\node[inner sep=0pt] (trank) at (5,3)
    {\includegraphics[width=.5in]{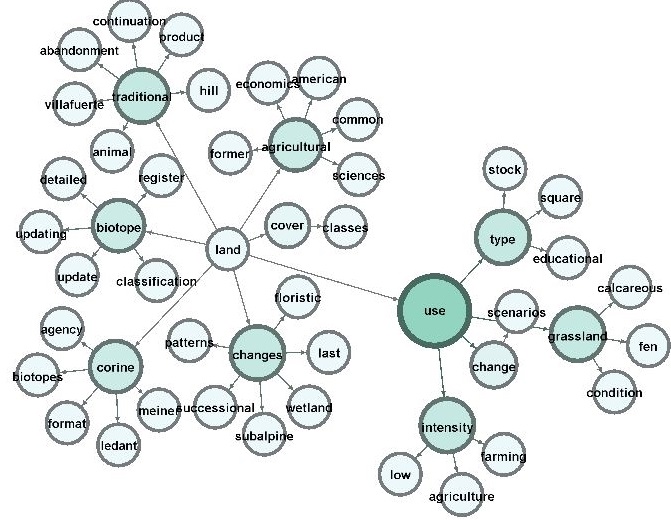}};
\node at (5,2.3) (b) {\scriptsize Text graph};
\node at (5,2) (b) {\scriptsize (Term co-occurrence)};
\draw[->,thick] (prep) -- (trank);

\node[inner sep=0pt] (prank) at (5,1)
    {\includegraphics[width=.5in]{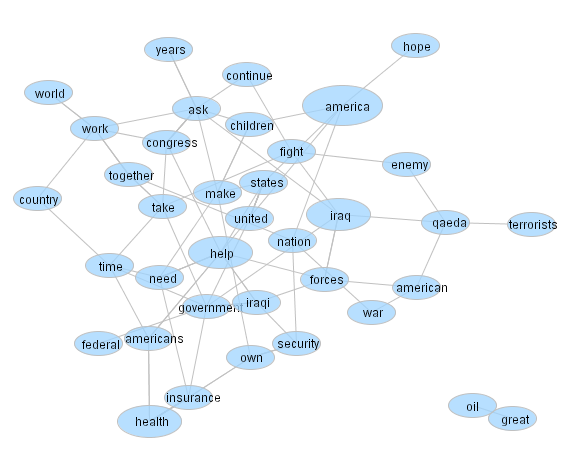}};
\node at (5,0.3) (b) {\scriptsize Text graph};
\node at (5,0) (b) {\scriptsize (POS dependence)};
\draw[->,thick] (prep) -- (prank);

\node[inner sep=0pt] (trc) at (7,3)
    {\includegraphics[width=.45in]{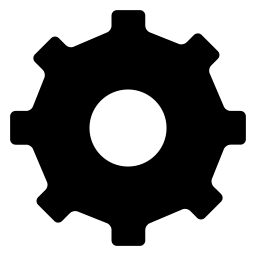}};
\node at (7,2.3) (b) {\scriptsize TextRank};
\node at (7,2) (b) {\scriptsize calculation};
\draw[->,thick] (trank) -- (trc);

\node[inner sep=0pt] (prc) at (7,1)
    {\includegraphics[width=.45in]{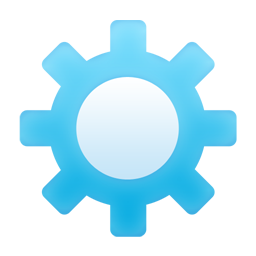}};
\node at (7,0.3) (b) {\scriptsize POSRank};
\node at (7,0) (b) {\scriptsize calculation};
\draw[->,thick] (prank) -- (prc);
\node[inner sep=0pt] (ranking) at (9,2)
    {\includegraphics[width=.5in]{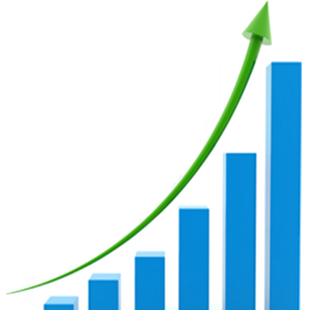}};
\node at (9,1) (b) {\scriptsize Term ranking};
\draw[->,thick] (trc) -- (ranking);
\draw[->,thick] (prc) -- (ranking);
\node[inner sep=0pt] (ranked) at (11,2)
    {\includegraphics[width=.5in]{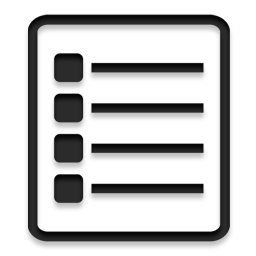}};
\node[inner sep=0pt] (ranked2) at (11.8,2)
    {\includegraphics[width=.25in]{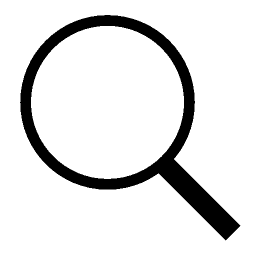}};
\node at (11,1) (b) {\scriptsize Ranked search};
\node at (11,0.7) (b) {\scriptsize terms (\bf{output})};

\draw[->,thick] (ranking) -- (ranked);
\end{pgfonlayer}
\end{tikzpicture}
\vspace{-.3cm}
\caption{Schematic diagram of the proposed technique--STRICT}
\label{fig:sysdiag}
\vspace{-.5cm}
\end{figure*}
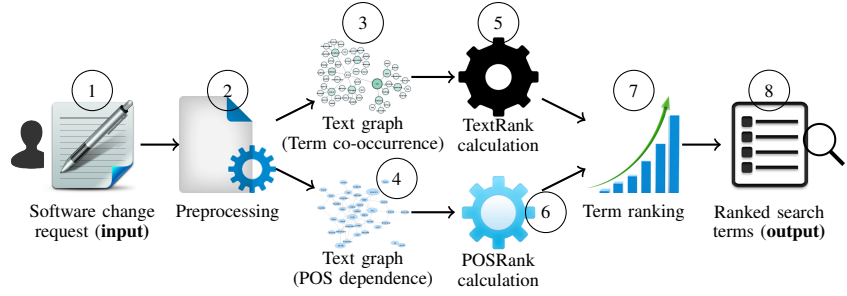

In information retrieval (IR), natural language text is often transformed into a graph where unique words are denoted as vertices, and meaningful relations among those words are represented as the edges \cite{rada}.
Such relation can be statistical, syntactic or semantic in nature \cite{blanco}.
In our research, we represent a software change request as a text graph where we consider both statistical and syntactic relations among words as the edges in the graph. 
To capture statistical relation, we consider co-occurrence of the words within a fixed window (\eg\ \emph{window size = 2}) across all sentences from the request.
For example, Fig. \ref{fig:tgraph}-(a) shows the text graph for the showcase change request (\ie\ Table \ref{table:ctask}) based on co-occurrence relationships among the words within a window of two.
In order to capture syntactic relation, we consider grammatical modification of words from the request using Jespersen's Rank Theory \cite{jespersen}. 
According to \citeauthor{jespersen}, words belonging to different parts of speech from a sentence can be provided with three
major ranks-- primary (\ie\ nouns), secondary (\ie\ verbs, adjectives) and tertiary (\ie\ adverbs)-- where a word from a higher rank modifies another word from the same or lower ranks.
We thus encode such modification relations into edges in the text graph. Fig. \ref{fig:tgraph}-(b) shows the text graph of the example change request (\ie\ Listing 1) based on grammatical dependencies among the words.
Once text graphs are developed, we apply two adapted versions of the popular algorithm by \citeauthor{pagerank} (for web link analysis)--\emph{PageRank} \cite{pagerank}--for term weighting.
In particular, we identify the most important words from the text graphs by exploiting their topological properties (\eg\ connectivity) in the graphs.

\section{STRICT: Proposed Technique for Search Term Identification from a Change Request}
\label{sec:proposed}
Given that appropriate search term identification is a major challenge for the developers and existing studies are limited in certain contexts, we introduce a novel IR-based technique.
Fig. \ref{fig:sysdiag} shows the schematic diagram of our proposed technique--STRICT--for search term identification and suggestion from a change request. 
We first turn a software change request into two text graphs (\eg\ Fig. \ref{fig:tgraph}) based on word co-occurrence and grammatical modification of words.
Then, we employ two graph-based term-weighting algorithms--\emph{TextRank} and \emph{POSRank}--on those graphs, estimate term weights, and identify the suitable search terms for the change task. 
In this section, we discuss different steps of our technique as follows:

\subsection{Data Collection}\label{sec:datacoll}
Our technique accepts user-provided texts of a change request as the input (\ie\ Step 1, Fig. \ref{fig:sysdiag}), and returns a ranked list of search terms as the output (\ie\ Step 8, Fig. \ref{fig:sysdiag}). 
We collect change requests from two popular bug tracking systems--\emph{BugZilla} and \emph{JIRA}.
Each change request is submitted as a semi-structured report written using natural language texts, and it contains several fields such as \emph{Issue ID} (\eg\ 303705), \emph{Product} (\eg\ JDT), \emph{Component} (\eg\ UI), \emph{Title} and \emph{Description}. 
We extract the last two fields from each report for analysis, as was also done by literature \cite{kevic}.
\emph{Title} summarizes a requested change task whereas \emph{Description} contains the user's  detailed explanation of the task in natural language texts.

\subsection{Text Preprocessing}\label{sec:preprocess}
We analyze \emph{Title} and \emph{Description} of a software change request, and perform several preprocessing steps on them (\ie\ Step 2, Fig. \ref{fig:sysdiag}). 
We consider \emph{sentence} as a logical unit for the change request texts, and collect each of the sentences from both fields. 
Then we perform standard natural language preprocessing (\ie\ stop word removal and splitting of dotted terms) on each of the sentences, and extract the candidate search terms.
A dotted term (\eg\ \texttt{org.eclipse.ui.part}) often contains multiple technical artifacts, and splitting helps one to analyze each of them in isolation \cite{splitting}. 
We also split each camel case (\eg\ \texttt{createPartControl}) term into simpler terms (\ie\ \texttt{create}, \texttt{Part} and \texttt{Control}) and keep both term types for our analysis \cite{splitting,refoqus}.
It should be noted that we avoid term stemming (\ie\ extracting root form of a given term) since it degrades the performance of our technique, as was also reported by \citet{kevic}.

\subsection{Text Graph Development}\label{sec:tokengraph}
\textbf{Using Word Co-occurrence:} After preprocessing steps, we get a list of sentences from each change request where each of the sentences contains an ordered list of candidate search terms. 
We then use those sentences to develop a \emph{text graph} (\eg\ Fig. \ref{fig:tgraph}-(a)) for the change request (Step 3, Fig. \ref{fig:sysdiag}).
In the text graph, each unique term is represented as a vertex and the \emph{co-occurrence} of terms in the sentence is denoted as the edges among the vertices. 
The underlying idea is that all the terms that co-occur in the text within a fixed window have some relationships or dependencies \cite{rada, blanco}.
For example, if we consider the sentence-- \emph{``Custom search results not shown hierarchically in the java search results view"}--from the example request texts (\ie\ Table \ref{table:ctask}), the preprocessed version
forms an ordered list of terms-- \emph{``Custom search hierarchically java search view."} 
Please note that the transformed sentence contains several phrases such as \emph{``custom search"} and \emph{``search view"}, and the terms in those phrases are semantically dependent on each other for comprehensive meaning.
Use of term co-occurrence captures such dependencies in a statistical sense.
We thus consider a \emph{sliding window} of \emph{window size=2} (as recommended by \citet{rada}) as a semantic unit of words, and derive the following relationships: 
\emph{Custom}$\longleftrightarrow$\emph{search}, \emph{search}$\longleftrightarrow$\emph{hierarchically}, \emph{hierarchically}$\longleftrightarrow$\emph{java}, \emph{java}$\longleftrightarrow$\emph{search} and \emph{search}$\longleftrightarrow$\emph{view}. 
Then such relationships are encoded into the connecting edges between the corresponding vertices in the text graph (\ie\ Fig. \ref{fig:tgraph}-(a)). 

\textbf{Using POS Dependence:} Term co-occurrence models relationship between terms statistically which might not be always effective for determining term-weight (\ie\ term's importance). We thus apply another type of relationship-- syntactic dependencies-- among the terms based on grammatical modification.   
According to Jespersen's Rank Theory \cite{jespersen}, words from a sentence can be provided with three major ranks-- primary (\ie\ nouns), secondary (\ie\ verbs, adjectives), and tertiary (\ie\ adverbs).
\citet{jespersen} suggests that a word from a higher rank generally defines (\ie\ modifies) another word from the same or lower ranks in a sentence.
Thus, a noun can modify only another noun whereas a verb can modify another noun, verb or adjective but not an adverb.
We consider this principle of grammatical modification of words, and represent such dependencies as directed edges in the text graph (\ie\ Step 4, Fig. \ref{fig:sysdiag}).
We first annotate each of the sentences from a change request using Stanford POS tagger \cite{postagger}, and organize the annotated words into various ranks.
For instance, the example statement---\emph{``element reported plain flat element hierarchical java search view"}--can be organized into two ranks--\emph{primary} (\emph{``search", ``view", ``java", ``element"}), and \emph{secondary} (\emph{``plain", ``flat", ``hierarchical"}, and \emph{``reported"}). 
We derive the following relationships based on their grammatical modifications-- \emph{search}$\longleftrightarrow$\emph{view}, \emph{view}$\longleftrightarrow$\emph{java}, \emph{java}$\longleftrightarrow$\emph{element},
\emph{reported}$\longrightarrow$\emph{search},
\emph{reported}$\longrightarrow$\emph{view},
\emph{reported}$\longrightarrow$\emph{java},
\emph{reported}$\longrightarrow$\emph{element},
\emph{reported}$\longrightarrow$\emph{plain},
\emph{reported}$\longrightarrow$\emph{flat},
\emph{reported}$\longrightarrow$\emph{hierarchical},
and, then encode them into connecting edges in the text graph (\ie\ Fig. \ref{fig:tgraph}-(b)).

\subsection{TextRank (TR) Calculation}
\label{sec:textrank}
Once text graph (\ie\ using co-occurrence) for a change request is developed, we consider it as a regular connected network, and apply a popular graph-based algorithm--PageRank \cite{pagerank}--for ranking its nodes (\ie\ terms) (Step 5, Fig. \ref{fig:sysdiag}). PageRank was originally proposed by \citeauthor{pagerank} for web link analysis, and the algorithm exploits topological properties of a graph to estimate the weight (\ie\ importance) of each of the vertices. 
TextRank is an adaptation of PageRank for text graph. It
analyzes the connectivity (\ie\ connected neighbours and their weights) of each term $v_i$ in the graph recursively, and then calculates the term's weight, $TR(v_{i})$, as follows:
\begin{equation*}\label{eq:textrank}
\setlength\abovedisplayskip{0pt}
\setlength\belowdisplayskip{0pt}
TR(v_{i})=(1-\phi)+\phi\sum_{j\epsilon V(v_{i})}\frac{TR(v_{j})}{|V(v_{j})|}~~ (0 \le \phi \le1)
\end{equation*}
\noindent
Here,  $V(v_{j})$ and $\phi$ denote node list connected to $v_i$ and \emph{dumping factor} respectively. In the text graph (\eg\ Fig. \ref{fig:tgraph}-(a)), co-occurrences among terms are represented as bi-directional connecting edges between the nodes.
In the context of web surfing, dumping factor, $\phi$, is considered as the probability of randomly choosing a web page by the surfer, and $1-\phi$ is the probability of jumping off that page. \citet{rada} use a heuristic value of $\phi=0.85$ for natural language texts in the context of keyword extraction, and we also use the same value for our TextRank calculation. 
We initialize each of the terms in the graph with a default value of 0.25,
and run an iterative version of the algorithm \cite{pagerank}. It should be noted that the initial value of a term does not affect its final score \cite{rada}.
The computation iterates until the scores of all the terms converge below a certain threshold or it reaches the maximum iteration limit (\ie\ 100 as suggested by \citet{blanco}). As \citet{rada} suggest, we use a heuristic threshold of 0.0001 for the convergence checking.

TextRank applies the underlying mechanism of a recommendation (\ie\ voting) system, where a term (\eg\ \emph{``Custom"}) recommends (\ie\ votes) another term (\eg\ \emph{``search"}) if the second term complements the semantics of the first term in any way (\eg\ \emph{``Custom search"}) \cite{rada}. 
The algorithm captures recommendation for a term by analyzing its connected edges (\ie\ both incoming and outgoing) in the text graph (\eg\ Fig. \ref{fig:tgraph}-(a)) with other terms both in local (\ie\ same sentence) and global (\ie\ entire task description) contexts, and thus determines importance of that term.
Once computation is over, each of the terms in the graph is found with a final score which is considered to be the weight or importance of that term within the user provided texts from the change request. 
 
\subsection{POSRank (POSR) Calculation}\label{sec:posrank}
While TextRank operates on a text graph based on word co-occurrence (\eg\ Fig. \ref{fig:tgraph}-(a)), POSRank determines term-weight by operating on the text graph (\eg\ Fig. \ref{fig:tgraph}-(b)) that represents the grammatical dependencies among words as the connecting edges (Step 6, Fig. \ref{fig:sysdiag}). 
POSRank is another adaptation of PageRank algorithm \cite{pagerank} for natural language texts. 
Similar to TextRank, it also analyzes connectivity of each term in the graph but considers the links according to their directions. Incoming links and outgoing links of the term are treated differently.  Incoming links represent votes cast for the term by other terms and vice versa. 
Thus, POSRank $POSR(v_{i})$ of each term $v_i$ is calculated as follows:
\begin{equation*}\label{eq:posrank}
\setlength\abovedisplayskip{0pt}
\setlength\belowdisplayskip{0pt}
POSR(v_{i})=(1-\phi)+\phi\sum_{j\epsilon In(v_{i})}\frac{POSR(v_{j})}{|Out(v_{j})|}~~ (0 \le \phi \le1)
\end{equation*}
\noindent
Here $In(v_i)$ and $Out(v_i)$ denote the node lists that are connected to node $v_i$ through incoming and outgoing links respectively.
Since the underlying mechanism of PageRank-based algorithms is recommendation (\ie\ votes) from other nodes of the graph, POSRank also determines the weight (\ie\ importance) of a term by capturing and analyzing the weights of the incoming links recursively. 
It should be noted that not only frequent votes but also the votes from other high scored nodes from the graph are essential for a node (\ie\ term) to be highly scored (\ie\ important).
Given the similar topological properties of the text graph using grammatical modifications (\ie\ Fig. \ref{fig:tgraph}-(b)),
we apply the same settings-- damping factor ($\phi$), iteration count, initial score, and convergence threshold--of TextRank (Section \ref{sec:textrank}) for POSRank calculation as well.
\setlength{\intextsep}{0pt}
\begin{algorithm}[h]
\caption{Search Term Identification using IR Methods}
\label{algo}
\begin{algorithmic}[1]
\Procedure{STRICT}{$CR$}\Comment{$CR$: change request}
\State $L \gets$ \{\}\Comment{list of search terms}
\LineComment{collecting task details from the change request}
\State $T \gets$ collectTitle($CR$)
\State $D \gets$ collectDescription($CR$)
\State $TD \gets$preprocess(combine($T,D$))
\LineComment{developing text graphs from the task details}
\State $G_{coc} \gets$ developTGUsingCo-occurrence($TD$)
\State $G_{pos} \gets$ developTGUsingPOS-dependence($TD$)
\LineComment{calculating TextRank and POSRank}
\State $TR \gets$ calculateTR($G_{coc}$)
\State $TR_{norm} \gets$ normalize(sortByValue($TR$))
\State $POSR \gets$ calculatePOSR($G_{pos}$)
\State $POSR_{norm} \gets$ normalize(sortByValue($POSR$))
\LineComment{calculating additional weights for title terms}
\State $TT \gets$ assignUniformWeight($T$)
\LineComment{determining term importance}
\State $CST \gets$ getUniqueTerms($TD$)
\For{CandidateSearchTerm $CST_i$ $\in$ $CST$}
\State $S_{TR} \gets TR_{norm}[CST_i]$ 
\State $S_{POSR} \gets POSR_{norm}[CST_i]$
 \State $S_{TT} \gets TT[CST_i]$
\LineComment{calculating final term-weight}
\State $S[{CST_i}] \gets S_{TR} +  S_{POSR} + S_{TT}$
\EndFor
\LineComment{ranking and then returning Top-K search terms}
\State $SST \gets$ sortByFinalTermWeight($S$)
\State $L \gets$ getTopKSearchTerms($SST$)
\State \textbf{return} $L$
\EndProcedure
\end{algorithmic}
\end{algorithm}

\vspace{-.3cm}
\subsection{Search Term Ranking and Selection}\label{sec:ranking}
Fig. \ref{fig:sysdiag} shows the schematic diagram and Algorithms \ref{algo}, \ref{algo2} show the pseudo code of our proposed technique--STRICT--for search term identification from a change request.
We first collect \emph{Title} and \emph{Description} of the request submitted by the user, combine them to prepare a complete request text, and then perform standard natural language preprocessing (Lines 3--6, Algorithm \ref{algo}).
Then we develop two text graphs based on co-occurrences and grammatical dependencies among the terms from the preprocessed text (Lines 7--9, Algorithm \ref{algo}). 
The goal is to identify the most important terms that could be used as a search query for concept location. 
We then analyze the topological properties of  both graphs, and determine TextRank and POSRank for each of the terms from the request (Lines 10--14, Algorithm \ref{algo}, Steps 5, 6, Fig. \ref{fig:sysdiag}).

\begin{algorithm}[h]
\caption{Candidate Score Normalization Algorithm}
\label{algo2}
\begin{algorithmic}[1]
\Procedure{NORMALIZE}{$R$}
\LineComment{$R$: candidates search terms sorted by scores}
\For{CandidateSearchTerm $t$ $\in$ $R.keys$}
\State $R[t] \gets 1-\frac{position(t)}{size(R)}$
\EndFor
\State \textbf{return} $R$ 
\EndProcedure
\end{algorithmic}
\end{algorithm}
\setlength{\textfloatsep}{0pt}
Since TextRank and POSRank estimate term importance from different perspectives, we normalize both scores for each of the candidate terms using Algorithm \ref{algo2}.  
Normalization step can reduce the potential bias of any specific score component.  
In particular, we sort the candidate terms based on TextRank or POSRank, and derive a score between 0 and 1 for each term by using its position in the sorted list. 
Such normalization technique is common in the relevant literature \cite{context,surfclipse}, and often called as \emph{degree of interest} \cite{context}.  
Given that title of the request often contains good quality search terms, we assign the highest degree of interest to the candidate terms found in the \emph{Title} field (Lines 15--16).    
Thus, each of the candidate terms gets three normalized scores from three types of ranking-- TextRank, POSRank and title-based heuristic. Please note that such scores for the candidates
are calculated by recursively analyzing the votes (\ie\ connections) from their surrounding terms in the text graphs.
Now, we iterate through the unique candidate search terms, and add up those three scores for each of the candidate terms (Lines 17--25, Algorithm \ref{algo}).
Then we rank the candidates based on their final accumulated scores, and select the Top-K ($K=10$) candidates as the search terms for the change request (Lines 26--30, Algorithm \ref{algo}, Step 7, 8, Fig. \ref{fig:sysdiag}).
We also expand the camel case search terms into simpler forms, and keep both forms in the search query \cite{splitting}.

\textbf{Example:} 
Table \ref{table:example} shows a working example of our technique--STRICT--for the showcase change request in Table \ref{table:ctask}. 
Our technique suggests the following Top-5 terms--\emph{`element', `IResource', `Provider', `Level'} and \emph{`Tree'}--that returns the first correct result at the Top-1 position.
We first calculate TextRank and POSRank for each of the candidate terms from the text graphs--Fig. \ref{fig:tgraph}-(a) and Fig. \ref{fig:tgraph}-(b)--respectively by carefully analyzing their topological properties (Sections \ref{sec:textrank}, \ref{sec:posrank}).
Table \ref{table:example} shows the normalized scores for the top terms. We see that these terms are highly connected in the text graphs, \ie\ frequently voted by other important terms across the request text. Our algorithms translate such connectivity into meaningful equivalent scores, and identify the top-scored terms carefully.
Please note that none of these terms comes from the title of the request which is often copied and pasted by the developers as the initial query.
Besides, the title returns the first correct result at the 559$^{th}$ position. Thus, our suggested terms can significantly reduce the query reformulation efforts spent by the developers, and can provide them with a meaningful starting point for concept location. 
\begin{table}
\centering
\caption{STRICT: A Working Example}\label{table:example}
\vspace{-.1cm}
\resizebox{2.3in}{!}{%
\begin{threeparttable}
\begin{tabular}{l|c|c|c|c}
\hline
\textbf{Search Term} & \textbf{TR} & \textbf{POSR}& \textbf{TH}  & \textbf{Score}\\
\hline
\hline
\emph{element} & 0.98 & 0.98 & 0 & 1.96\\
\hline
\emph{IResource} & 0.93 & 1.00 &  0 & 1.93 \\
\hline
\emph{Provider} & 0.68 & 0.95 &  0 & 1.63 \\
\hline
\emph{Level} & 0.83 & 0.80 & 0 & 1.63  \\
\hline
\emph{Tree} & 0.78 & 0.75 & 0 & 1.53  \\
\hline
\end{tabular}
\centering
 \textbf{TR:} TextRank, \textbf{POSR:} POSRank, \textbf{TH:} Title Heuristic
\end{threeparttable}
}
\vspace{-.4cm}
\end{table}

\begin{table}[!t]
\centering
\caption{Experimental Dataset}\label{table:dataset}
\vspace{-.2cm}
\resizebox{3.5in}{!}{%
\begin{threeparttable}
\begin{tabular}{l|c|c||l|c|c}
\hline
\textbf{System} & \textbf{\#MD} & \textbf{\#CR} & \textbf{System}& \textbf{\#MD} & \textbf{\#CR}\\
\hline
\hline
\texttt{eclipse.jdt.core-4.7.0}& 64K & 292 & \texttt{ecf-170.170}& 21K & 262  \\
\hline
\texttt{eclipse.jdt.debug-4.6.0}& 16K & 63 & \texttt{log4j-1.2.17}& 3K & 60  \\
\hline
\texttt{eclipse.jdt.ui-4.7.0}& 57K & 419 & \texttt{sling-0.1.10}& 30K & 76  \\
\hline
\texttt{eclipse.pde.ui-4.7.0}& 32K & 329 & \texttt{tomcat-7.0.70}& 24K & 438 \\
\hline
\end{tabular}
\centering
\textbf{MD:} \textbf{M}ethod \textbf{D}efinitions, \textbf{CR:} \textbf{C}hange \textbf{R}equests selected for experiments
\end{threeparttable}
}
\end{table}

\section{Experiment}
\label{sec:experiment}
Given two methods--\emph{pre-retrieval} and \emph{post-retrieval}--in the literature \cite{qeffect,qquality}, we chose post-retrieval method for the evaluation and validation of our suggested search queries. This method directly evaluates results returned by a query, and thus is more reliable than the other.
Besides, past relevant studies \cite{twkraft,trconfig,refoqus} also adopted this method for evaluation and validation.
We evaluate our technique--STRICT--using 1,939 software change requests with four appropriate performance metrics,  and compare with two state-of-the-art techniques \cite{kevic,rocchio}.
Thus, we answer four research questions as follows:
\begin{itemize}[noitemsep]
\item \textbf{RQ$\mathbf{_1}$:} Are our suggested queries significantly better in term of effectiveness than the baseline search queries from the software change requests?
\item \textbf{RQ$\mathbf{_2}$:} How do our suggested search queries actually perform in retrieving correct/relevant results?     
\item \textbf{RQ$\mathbf{_3}$:} How effective are the information retrieval algorithms--\emph{TextRank} and \emph{POSRank}--in identifying good quality search terms from a change request?
\item \textbf{RQ$\mathbf{_4}$:} Can our proposed technique--STRICT-- outperform the state-of-the-art techniques in identifying good quality search terms from a change request? 
\end{itemize}

\vspace{-.2cm}
\subsection{Experimental Dataset}\label{sec:dataset}
\textbf{Data Collection:} We collect 1,939 software change requests from eight Java-based subject systems (\ie\ five \emph{Eclipse} systems and three \emph{Apache} systems) for our experiments.
Table \ref{table:dataset} shows the details of our selected systems.  
Each of these change requests was marked as \emph{RESOLVED}, and we follow a careful approach in their selection. 
We first collect the \emph{RESOLVED} change requests from \emph{BugZilla} and \emph{JIRA} bug repositories of those systems, and then map them to the commit history of their corresponding projects at GitHub and SVN.
We analyze the commit messages from each project, and look for specific Issue ID (\ie\ identifier of a change task) in each of those messages \cite{bugid}. 
Then, we include any change request in the experimental dataset only if there exists a corresponding commit in the collected commit history. 
Such evaluation approach is regularly adopted by the relevant literature \cite{kevic,refoqus,twkraft,buglocator}, and we also follow the same. In order to ensure a fair evaluation, we also discard the change requests from our dataset for which no source code files (\ie\ Java classes) were changed or the relevant code files were missing from the system.

\textbf{Goldset Development:}
In GitHub and SVN, we note that any commit that either solves a software bug or implements a feature request generally mentions the corresponding Issue ID in the very title of the commit.
We identify such commits from the commit history of each of the selected projects using suitable regular expressions, and select them for our experiments \cite{bugid}. Then, we collect the \emph{changeset} (\ie\ list of changed files) for each of those commit operations, and develop solution set (\ie\ \emph{goldset}) for the corresponding change tasks. Thus, for experiments, we collect not only the actual change requests from the reputed subject systems but also their solutions which were applied in practice by the developers \cite{specificity}. We use several utility commands such as \emph{git, clone, rev-list} and \emph{log} on GitHub and SVN consoles for collecting those information. 

\textbf{Replication:} All experimental data and supporting materials are hosted online \cite{strict} for replication or third-part reuse.

\begin{table*}[!t]
\centering
\caption{Effectiveness Details of STRICT Query vs. Baseline (Title)}\label{table:effectiveness}
\vspace{-.2cm}
\resizebox{7.2in}{!}{%
\begin{threeparttable}
\begin{tabular}{l|c|c|c|c|c|c|c|c||c|c|c|c|c|c|c||c}
\hline
\multirow{2}{*}{\textbf{System}} & \multirow{2}{*}{\textbf{\#Queries}} &\multicolumn{7}{c||}{\textbf{Improvement}} & \multicolumn{7}{c||}{\textbf{Worsening}} & \textbf{Preserving}   \\
\hhline{~~---------------}
& & \#Improved & Mean & Q1 & Q2 & Q3 & Min. & Max. & \#Worsened & Mean & Q1 & Q2 & Q3 & Min. & Max. & \#Preserved\\
\hline
\hline
\textbf{ecf} & 262 & \textbf{145 (55.34\%)} & 220 & \textbf{7} & 27 &  140 & 1 & 3,450 & 100 (38.17\%) & 648 & 65 & 195 & 815 &  2 & 6,417 & \textbf{17 (6.49\%)} \\
\hline
\textbf{jdt.core}  & 292 & \textbf{154 (52.74\%)} & 421 & \textbf{6} &  46 & 207 & 1 & 6,636 & 122 (41.78\%) & 1050 & 112 & 346 & 932 & 3 & 9,419 & \textbf{16 (5.48\%)}\\
\hline
\textbf{jdt.debug}  & 63 & \textbf{45 (71.43\%)} & 352 & \textbf{10} &  47 & 155 & 1 & 4,401 & 15 (23.81\%) & 682 & 66 & 195 & 1,008 & 9 & 2,942 & \textbf{3 (4.76\%)}\\
\hline
\textbf{jdt.ui}  & 419 & \textbf{229 (54.65\%)} & 282 & \textbf{7} &  26 & 108 & 1 & 9,851 & 179 (42.72\%) & 974 & 55 & 222 & 916 & 2 & 8,596 & \textbf{11 (2.63\%)}\\
\hline
\textbf{pde.ui}  & 329 & \textbf{169 (51.37\%)} & 163 & \textbf{10} &  34 & 125 & 1 & 2,746 & 128 (38.91\%) & 1033 & 87 & 330 & 1,272 & 7 & 7,602 & \textbf{32 (9.72\%)}\\
\hline
\textbf{log4j}  & 60 & \textbf{30 (50.00\%)} & 74 &  \textbf{7} &  13 & 73 & 1 & 788 & 20 (33.33\%) & 197 & 47 & 124 & 296 & 5 & 630 & \textbf{10 (16.67\%)}\\
\hline
\textbf{sling}  & 76 & \textbf{46 (60.53\%)} & 257 & \textbf{7} &  24 & 236 & 1 & 2,849 & 24 (31.58\%) & 501 & 25 & 141 & 564 & 7 & 2,904 & \textbf{6 (7.89\%)}\\
\hline
\textbf{tomcat70}  & 438 & \textbf{292 (66.67\%)} & 152 & \textbf{4} &  14 & 100 & 1 & 5,846 & 128 (29.22\%) & 555 & 70 & 229 & 675 & 3 & 5,918 & \textbf{18 (4.11\%)}\\
\hline
& Total=1,939 & \textbf{Avg = 57.84\%}  & &  &  &  &  &  & Avg = 34.94\%  &  &  &  &  &  &  & \textbf{Avg = 7.22\%}    \\
\hline
\end{tabular}
\centering
\textbf{jdt.core} = \texttt{eclipse.jdt.core}, \textbf{jdt.debug} = \texttt{eclipse.jdt.debug}, \textbf{jdt.ui} = \texttt{eclipse.jdt.ui}, \textbf{pde.ui} = \texttt{eclipse.pde.ui}, \textbf{Mean}=Mean rank of first relevant document in the search result, \textbf{Q$_i$}= Rank value for $i^{th}$ quartile of all result ranks
\end{threeparttable}
}
\vspace{-.7cm}
\end{table*}

\vspace{-.1cm}
\subsection{Search Engine}\label{sec:sengine}
We use a Vector Space Model (VSM) based search engine-- \emph{Apache Lucene} \cite{qeffect,refoqus}--to search files that were changed for implementing the change requests. 
Search engines generally index the files in a corpus prior to search. 
Since Lucene indexer is targeted for simple text documents (\eg\ news article) and the source code files in our selected projects contain items beyond regular texts, \ie\ code segments, we apply limited preprocessing on them.
In particular, we extract method bodies from each of the Java classes (\ie\ source files), and consider each method as an individual document in the corpus (Table \ref{table:dataset}). 
We remove all programming keywords and punctuation characters, and split all dotted and camel case tokens \cite{splitting}.
Please note that we avoid stemming of the tokens for aforementioned reasons as described in Section \ref{sec:preprocess} \cite{kevic}.
This preprocessing step transforms the source files into text like files, and the indexer becomes able to perform more effectively, especially in choosing meaningful index terms.  
Once a search is initiated using a query, the search engine collects relevant documents from the corpus using a \emph{Boolean Search Model}, and then applies a \emph{tf-idf} based scoring technique to return a ranked list of relevant documents.
As existing studies suggest \cite{kevic, topten}, we consider the Top-10 results from the search engine for calculating the performance of our suggested queries.

\subsection{Performance Metrics} \label{sec:pmetrics}
We choose four performance metrics for evaluation and validation of our suggested search queries.
These metrics are frequently used by the relevant literature \cite{refoqus,stacktrace,saha}, and thus are highly appropriate for our experiments as well.

\textbf{Effectiveness (E)}: It is a measure that approximates the developer's effort for locating a concept in the source code \cite{stacktrace}.
In short, the measure returns the rank of the first occurred file (\ie\ from the gold set) in the search result list. 
The lower the effectiveness value is, the more effective a query is.


\textbf{Mean Reciprocal Rank@K (MRR@K)}: Reciprocal rank@K refers to the multiplicative inverse of the rank of the first correctly returned changed file (\ie\ from gold set) within the Top-K results.
Mean Reciprocal Rank@K (MRR@K) averages such measures for all change requests.

\textbf{Mean Average Precision@K (MAP@K)}: Precision@K calculates precision at the occurrence of every single relevant result in the ranked list. Average Precision@K (AP@K) averages the precision@K for all relevant results in the list for a search query. 
Thus, Mean Average Precision@K is calculated from the mean of average precision@K for all queries. 

\textbf{Top-K Accuracy}: It refers to the percentage of the change requests for which at least one changed file is correctly returned within the Top-K results by the search queries.


\begin{table}
\centering
\caption{Comparison of STRICT'S Effectiveness with Baseline Queries}\label{table:baseline}
\vspace{-.2cm}
\resizebox{3.5in}{!}{%
\begin{threeparttable}
\begin{tabular}{l|c|c|c|c|c}
\hline
\textbf{Query Pairs} & \textbf{Improved} & \textbf{Worsened}  & \textbf{p-value} & \textbf{Preserved} & \textbf{MRD} \\
\hline
\hline
STRICT vs. Title & \textbf{57.84}\% &  34.94\% & *($<$0.001)  & \textbf{7.22}\% & -147\\
\hline
STRICT vs.   & \textbf{62.49}\% &   32.26\% & *($<$0.001)  & \textbf{5.25}\% & -201\\
\hhline{~~~~~~}
Title (10 keywords) & & & & & \\
\hline
STRICT vs. Description & \textbf{53.84}\% & 38.21\% & *($<$0.001) & \textbf{7.95}\% & -329 \\
\hline
STRICT vs. & \textbf{52.36}\%  &  39.94\% & *($<$0.001) & \textbf{7.70}\% & -265\\
\hhline{~~~~~~}
(Title + Description) & & & & & \\
\hline
\end{tabular}
\centering
\textbf{*} = Statistically significant difference between improvement and worsening, \textbf{MRD} = Mean Rank Difference between STRICT and baseline queries 
\end{threeparttable}
}
\end{table}

\subsection{Evaluation of STRICT}\label{sec:evaluation}
We conduct experiments using 1,939 change requests from eight subject systems (Table \ref{table:dataset}) where the above four metrics (Section \ref{sec:pmetrics}) are applied to performance evaluation.
We run each of our suggested queries with a search engine (Section \ref{sec:sengine}), check their results against the collected gold set files (Section \ref{sec:dataset}), and then compare with baseline queries from those requests. 
In this section, we discuss our evaluation details and answer \textbf{RQ$\mathbf{_1}$}, \textbf{RQ$\mathbf{_2}$}, and \textbf{RQ$\mathbf{_3}$} as follows.

\textbf{Answering \textbf{RQ$\mathbf{_1}$}--Comparison with Baseline Queries}:
Developers often copy and paste the text from a software change request on an ad-hoc basis for searching the source code that needs to be changed. Thus, change requests can be considered as the baseline queries for evaluation or validation of our suggested queries \cite{kevic,refoqus,twkraft}. 
We consider three types of baseline queries from the fields of each request--\emph{Title, Description}, and \emph{Title}+\emph{Description}, run them against our search engine, and then collect the rank of the first correctly returned result for each of those queries.
We also collect similar ranks returned by our suggested queries for each of those change requests, and compare with that of baseline queries.
Tables \ref{table:baseline} and \ref{table:effectiveness} report our comparative analysis.
From Table \ref{table:baseline}, we see that our suggested queries are more effective than the baseline queries.
Our queries provide better ranks than the baseline queries for 52\% -62\% of the cases while preserving the quality of 5\%--8\% which are promising according to the relevant literature \cite{refoqus,trconfig}.  
They provide relatively lower ranks than baseline for 35\% of the requests on average.
We also compare the query improvement ratios and worsening ratios by our technique using statistical significance test-- \emph{Mann-Whitney U (MWU) test}, and found that our improvements are significantly higher (\ie\ all $p-values<0.05$, Table \ref{table:baseline}) with each type of baseline queries. 
For example, when compared with \emph{Title} from the request, our queries provide 58\%  improvement ratio on average with 35\% worsening ratio.
More interestingly, when first 10 keywords (\ie\ average size of our request titles) are chosen from \emph{Title} like we suggest Top-10 search terms, the improvement ratio 
(\ie\ 62\%) is almost \emph{twice} the worsening ratio (\ie\ 32\%) by our suggested queries, which demonstrates the high potential of our technique.
Mean Rank Difference (MRD) between our provided queries and baseline queries is negative (\eg\ -329) which suggests that our queries generally return the goldset files at upper positions of the result list than the baseline queries. Table \ref{table:effectiveness} provides further details on the \emph{effectiveness} of our queries. We see that our queries provide better ranks (\ie\ 50\%--71\%) than the baseline ranks of \emph{Title} for each of the eight subject systems, which is promising. More importantly, the 25\% quantile ($Q_1$) of the improved ranks for each of the subject systems is $Q_1\le 10$.
That means, for one-fourth of the improved cases from each system, STRICT provides the relevant results within the Top-10 positions.
When we consider all returned ranks (\ie\ improved, worsened and preserved) for those systems, we found that 15\%--33\% of our first correct ranks are within Top-10 positions whereas such statistic for the baseline queries is only 02\%--11\%, which demonstrates the potential of our technique. 

Earlier studies report significant benefit in including source code tokens in the search query during bug localization \cite{stacktrace} and feature location \cite{twkraft}.
We also thus compare with code-only baseline queries where each query comprises of camel case tokens extracted from the change request.    
We found that our queries provide better ranks than 63\% of the code-only baseline queries while returning lower ranks for only 29\%. 
We manually investigated, and found that 37\% of our baseline queries do not contain any source code tokens, which possibly explains such finding.
Thus, while source code tokens might make a good search query, their presence in the change request cannot be guaranteed and thus, queries based on them could also be limited.
On the contrary, we suggest relatively better queries from the careful analysis of any available textual information in the change request.  
We also manually investigate the worsened queries by STRICT and corresponding requests, and found two important observations. First, the extent of our rank worsening is relatively smaller than that of rank improvement. Second, most of those requests contain structural content (\eg\ stack traces), and 
STRICT could have performed even better if such structures were properly incorporated in the text graphs.

Thus, to answer \textbf{RQ$\mathbf{_1}$}, our suggested queries are significantly more effective than baseline queries from the change requests. They provide better ranks than baseline queries for 52\%--62\% of the requests which is promising.

\begin{figure}[!t]
\centering
\includegraphics[width=2.6in]{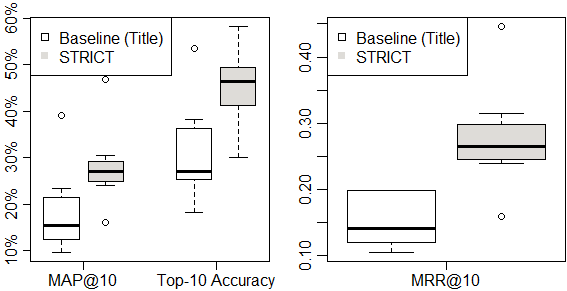}
\vspace{-.3cm}
\caption{Comparison of retrieval performance with baseline query (\ie\ title)}
\vspace{-.3cm}
\label{fig:performance}
\end{figure}

\begin{figure}[!t]
\centering
\includegraphics[width=1.8in]{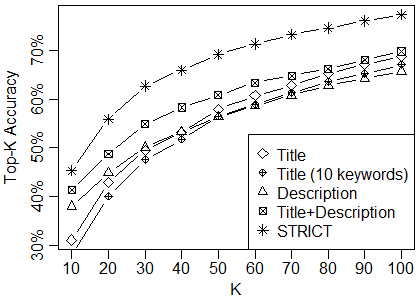}
\vspace{-.4cm}
\caption{Comparison of Top-K accuracy with all baseline queries}
\label{fig:topk}
\end{figure}

\textbf{Answering RQ$\mathbf{_2}$--Retrieval Performance:}
Although our suggested queries provide significantly better ranks than baseline queries in RQ$_1$, detailed retrieval performance could be another indicator of their quality.
We thus run each of our queries, analyze the Top-10 results (\ie\ as many existing studies do \cite{kevic,refoqus,pick}), and calculate Top-K accuracy, mean average precision@K and mean reciprocal rank@K of our technique.
Table \ref{table:performance}, Fig. \ref{fig:performance} and Fig. \ref{fig:topk} report our performance.
From Table \ref{table:performance}, we see that our queries perform correctly for 45\% of the change requests with 28\%  mean average precision@10 and 
a mean reciprocal rank@10 of 0.28. That is, one can get the relevant files (\ie\ that need to be changed) with our queries within the Top-4 results for 45\% of the times, which is promising. Such performance is also comparable to other existing findings from the literature \cite{giovanni,antoniol,twkraft}.
Although our retrieval performance is not much higher, our technique--STRICT--has more potential than the baseline queries for practical usage.
It achieves relatively higher performance with a limited number of search terms. For example, baseline queries--\emph{Description} and \emph{Title}+\emph{Description}--perform close (\ie\ 38\%--41\% Top-10 accuracy) to ours (\ie\ 45\% Top-10 accuracy). However, their performance is achieved at the cost of about 150 terms  in each query which is extremely difficult for any developer to reformulate or choose keywords from. On the other hand, 5--10 suggested terms from STRICT provide 40\%--45\% Top-10 accuracy which demonstrates their higher potential.
From the viewpoint of reformulation efforts, \emph{Title} can be considered as a potential baseline opponent, and we compare with this type of queries extensively.
We perform statistical significance test, and found that our performance is significantly higher than the baseline query--\emph{Title}--for all three performance metrics--mean average precision@10 (\ie\ $p-value: 0.010<$0.05), mean reciprocal rank@10 (\ie\ $p-value: 0.007<$0.05) and Top-10 accuracy (\ie\ $p-value: 0.007<$0.05). 
The box plots in Fig. \ref{fig:performance} also demonstrate that STRICT performs better than the baseline in each of the measures with higher medians.
We also investigate how Top-K values affect the retrieval performance of the queries where K takes up values from 10 to 100.
From Fig. \ref{fig:topk}, we see that STRICT provides a Top-100 accuracy of 77\% whereas such statistics for the baseline queries are 66\%--70\%.
More importantly, our accuracy measures always remained significantly higher than the baseline measures--\emph{Title} (\ie\ $p-value: 0.012<0.05$), \emph{Description} (\ie\ $p-value: 0.005<0.05$), and \emph{Title}+\emph{Description} (\ie\ $p-value: 0.031<0.05$)--for various K-values which demonstrate the actual strength of the proposed technique.

\begin{table}
\centering
\caption{Retrieval Performance of STRICT Queries}\label{table:performance}
\vspace{-.2cm}
\resizebox{3.5in}{!}{%
\begin{threeparttable}
\begin{tabular}{l|c|c|c|c}
\hline
\textbf{Query} & \textbf{\#Keywords} & \textbf{MAP@10} & \textbf{MRR@10} & \textbf{Top-10 Accuracy} \\
\hline
\hline
Title &  10 & \textbf{18.40}\%  &  \textbf{0.18} & \textbf{31.14}\% \\
\hline
Title (10 keywords)&  09 & 16.63\%  &  0.16 & 27.95\% \\
\hline
Description & 146 & 22.60\% & 0.23 & 38.00\% \\
\hline
Title + Description& 156  & \textbf{24.54}\%  &  \textbf{0.24} & \textbf{41.40}\% \\
\hline
\hline
Title + Description& 15  & 19.98\%  &  0.20 & 29.51\%\\
(Code tokens only) &  & & & \\
\hline
Title + Description& 135  & 18.87\%  &  0.18 & 34.03\%\\
(Without code tokens) &  & & & \\
\hline
\hline
STRICT$_{T3}$ & 03 & 17.93\%  &  0.17 & 31.32\% \\
\hline
STRICT$_{T5}$ & 05 & 24.38\%  &  0.24 & 39.90\%\\
\hline
STRICT & 10 & \textbf{28.09}\%*  &  \textbf{0.28}* & \textbf{45.34}\%* \\
\hline
\end{tabular}
\centering
\textbf{*} = Statistically significant difference between suggested and baseline (\ie\ title) queries
\end{threeparttable}
}
\vspace{-.5cm}
\end{table}

We also investigate the retrieval performance of baseline queries containing source code tokens only and without any code tokens. From Table \ref{table:performance}, we see that combination of both token types produced better quality search queries (41\% Top-10 accuracy) than any single type alone (\ie\ 30\%--34\% Top-10 accuracy). However, our technique provides even better queries in terms of all retrieval performance metrics.

Thus, to answer \textbf{RQ$\mathbf{_2}$}, our suggested queries provide 45\% Top-10 accuracy and 77\% Top-100 accuracy with moderate mean average precision and mean reciprocal rank which are significantly higher than those of baseline queries (\eg\ title of a change request) and are also promising.

%
%

\textbf{Answering \textbf{RQ$\mathbf{_3}$}--Role of IR Algorithms:} We investigate how the two term-weighting techniques--TextRank and POSRank-- perform in identifying good quality search terms from a software change request.   
Table \ref{table:role} summarizes our investigation details.
We see that both TextRank and POSRank perform almost equally well in terms of all four performance metrics.  Our technique--STRICT-- provides about 45\% Top-10 accuracy with each term-weight considered which shows their potential.
However, their combination in STRICT provides marginal improvement in precision (\ie\ MAP@10) and reciprocal rank (\ie\ MRR@10), which possibly justifies our choice for the combination. Similar scenario is observed in the case of query effectiveness. Combination of TextRank and POSRank reduces worsening ratio and increases the improvement ratio to 58\% whereas the individual ratio is 55\% for each of them.


Thus, to answer \textbf{RQ$\mathbf{_3}$}, both TextRank and POSRank are found effective in identifying good quality search terms from a change request.
Each of them alone provides such effective search queries that can return better ranks than 55\% of the baseline queries on average, which is promising.

\begin{table}
\centering
\caption{Role of TextRank and POSRank in STRICT Queries}
\label{table:role}
\vspace{-.2cm}
\resizebox{3.3in}{!}{%
\begin{threeparttable}
\begin{tabular}{l|l|c||l|c}
\hline
& \multicolumn{2}{c||}{\textbf{Retrieval Performance}} & \multicolumn{2}{c}{\textbf{Effectiveness} } \\
\hhline{~----}
\textbf{Term weight} & \textbf{Metric} & \textbf{Value}  & \textbf{Metric} & \textbf{Value}\\
\hline
\hline
\multirow{3}{*}{TextRank} & Top-10 Accuracy & 46.02\% &  Improved & 54.85\%\\
\hhline{~----}
& MRR@10 & 0.26 & Worsened & 37.05\% \\
\hhline{~----}
& MAP@10 & 26.75\%  & Preserved & 8.10\% \\
\hline
\hline
\multirow{3}{*}{POSRank} & Top-10 Accuracy & 45.31\% &  Improved & 54.52\%\\
\hhline{~----}
& MRR@10 & 0.27 & Worsened & 37.13\% \\
\hhline{~----}
& MAP@10 & 27.63\%  & Preserved & 8.35\% \\
\hline
\hline
\multirow{3}{*}{} & Top-10 Accuracy & \textbf{45.34}\% &  Improved & \textbf{57.84}\%\\
\hhline{~----}
TextRank +& MRR@10 & \textbf{0.28} & Worsened & 34.94\% \\
\hhline{~----}
POSRank& MAP@10 & \textbf{28.09}\%  & Preserved & \textbf{7.22}\% \\
\hline
\end{tabular}
\centering
\end{threeparttable}
}
\end{table}

\subsection{Comparison with Existing Techniques}\label{sec:comparison}
\begin{figure}[!t]
\centering
\includegraphics[width=2.8in]{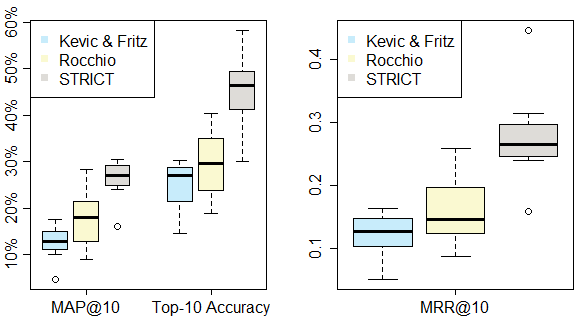}
\vspace{-.4cm}
\caption{Comparison with existing techniques}
\vspace{-.4cm}
\label{fig:compare}
\end{figure}

\begin{figure}[!t]
\centering
\includegraphics[width=2in]{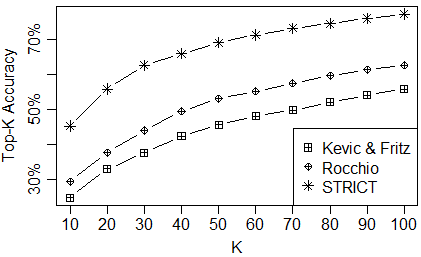}
\vspace{-.4cm}
\caption{Comparison of Top-K accuracy with existing techniques}
\label{fig:compare-topk}
\end{figure}

STRICT can be considered both as (1) a search term identification technique and (2) a search query reformulation technique. It not only identifies good quality search terms from a change request but also, in essence, reformulates the baseline queries by discarding low quality search terms.
We thus compare our technique with two relevant existing studies--\citet{kevic} and \citet{refoqus}--that identify search terms and reformulate search queries from software change requests respectively.   
To the best of our knowledge, \citet{kevic} is the only available study in the literature for search term identification, and Rocchio's expansion \cite{rocchio} is reported to be one of the best performing strategies for query reformulation by \citet{refoqus}.
Thus, these two can be considered as the state-of-the-art techniques.  We select them for comparison using our experimental dataset (Section \ref{sec:dataset}), and Table \ref{table:comparison}, Fig. \ref{fig:compare} and Fig. \ref{fig:compare-topk} report our comparative analyses as follows. 


\textbf{Answering RQ$\mathbf{_4}$--Comparison with the State-of-the-Art:}
\citet{kevic} use a heuristic-based regression model in identifying search terms from a change request where they consider frequency (\ie\ \emph{tf-idf}), location (\ie\ \emph{inSumAndBody, isInMiddle}) and notation (\ie\ \emph{isCamelCase}) of the terms from the request content. 
\citet{rocchio} expands a baseline query by collecting candidate terms from the top  $K=5$ source files returned by that query and then applying \emph{tf-idf} based term weighting to each of them \cite{refoqus}. 
We implement both methods using their provided settings (\eg\ metric weight), collect their search queries for the change requests, and then evaluate those queries using the same search engine they used (\ie\ Lucene). 


\begin{table}
\centering
\caption{Comparison with Existing Techniques}
\label{table:comparison}
\vspace{-.2cm}
\resizebox{3.5in}{!}{%
\begin{threeparttable}
\begin{tabular}{l|l|c||l|c}
\hline
& \multicolumn{2}{c||}{\textbf{Retrieval Performance}} & \multicolumn{2}{c}{\textbf{Effectiveness}}\\
\hhline{~----}
\textbf{Technique} & \textbf{Metric} & \textbf{Value}  & \textbf{Metric} & \textbf{Value}\\
\hline
\hline
\multirow{4}{*}{\citet{kevic}} & Top-10 & \multirow{2}{*}{24.90\%} &  Improved & \textbf{40.09}\%\\
\hhline{~~~--}
& Accuracy &  & Worsened & 53.95\% \\ 
\hhline{~----}
& MRR@10 & 0.12  & Preserved & 5.96\% \\
\hhline{~----}
 & MAP@10 & 12.45\% & MRD  & +101\\
\hline
\hline
\multirow{4}{*}{Rocchio's Method \cite{rocchio,refoqus}} & Top-10 &   \multirow{2}{*}{\textbf{29.48}\%}  &  Improved & 37.59\%\\
\hhline{~~~--}
& Accuracy &  & Worsened & 56.38\% \\
\hhline{~----}
& MRR@10 & 0.16  & Preserved & 6.03\% \\
\hhline{~----}
& MAP@10 & \textbf{17.66}\%  & MRD & +45\\
\hline
\hline
\multirow{2}{*}{} & Top-10 & \multirow{2}{*}{\textbf{45.34}\%*} &  Improved & \textbf{57.84}\%*\\
\hhline{~~~--}
STRICT& Accuracy  & & Worsened & 34.94\%* \\
\hhline{~----}
(Proposed approach) & MRR@10 & \textbf{0.28}*  &  Preserved & \textbf{7.22}\% \\
\hhline{~----}
 & MAP@10 & \textbf{28.09}\%*  & MRD & \textbf{-147}\\
\hline
\end{tabular}
\centering
\textbf{*} = Statistically significant difference between proposed and existing techniques
\end{threeparttable}
}
\vspace{-.4cm}
\end{table}

From Table \ref{table:comparison}, we see that our queries are more effective than the queries provided by \citet{kevic} or Rocchio's method \cite{rocchio}.
In the case of retrieval performance, the close competitor is Rocchio's method whereas \citeauthor{kevic} performs the better (of the two) in the case of query effectiveness.
They provide 29\% Top-10 accuracy and 40\% rank improvement (than baseline ranks) at best.   
On the contrary, our technique--STRICT--provides 45\% Top-10 accuracy with 28\% mean average precision@10 and a mean reciprocal rank@10 of 0.28 which are promising.
It also improves 58\% of 1,939 baseline queries which is highly promising according to relevant literature \cite{refoqus,trconfig}.
We also found that all of our reported measures--\emph{accuracy, precision, reciprocal rank, improvement} and \emph{worsening}--are significantly higher (\ie\ all $p-values<0.05$) than that of these two competing techniques. When the retrieval performance measures are plotted using box plots in Fig. \ref{fig:compare}, we also see that our performance is quite higher than the state-of-the-art with higher medians and lower variances. More interestingly, when we plot Top-K accuracy against various K-values in Fig. \ref{fig:compare-topk}, our accuracy always remained significantly higher than \citet{kevic} (\ie\ $p-value:<0.001$) and \citet{rocchio} (\ie\ $p-value:0.001<0.05$).
The state-of-the-art provides 63\% Top-100 accuracy whereas that measure for STRICT is 77\%, which clearly demonstrates the superiority of our technique. 
  
Thus, to answer \textbf{RQ$\mathbf{_4}$}, our technique--STRICT-- performs significantly better than the state-of-the-art in identifying good quality search terms from a given change request. Such terms not only provide better retrieval performance (\eg\ 77\% Top-100 accuracy) but also reformulate a baseline query more effectively (\eg\ 58\% improvement) than the state-of-the-art.

\section{Threats to Validity}\label{sec:threat}

Threats to internal validity relate to experimental errors and biases \cite{wordsim}. 
Re-implementation of the existing techniques is a possible source of such threat. 
Due to the lack of reliable or directly applicable prototypes,   
both existing techniques--\citet{kevic} and \citet{rocchio}--were  re-implemented.
However, these techniques are based on two different equations with clearly stated independent and dependent variables, and we implemented them carefully.
Besides, we ran them in our experiments multiple times, and compared with their best performance. 
Thus, such threat might be mitigated.

Threats to external validity relate to the generalizability of a technique \cite{saner2016masud}. So far, we experimented with only Java-based systems. However, given the simplicity in our project corpus creation (Section \ref{sec:sengine}), our technique can be easily replicated for subject systems using other programming languages.



The POS tagging might contain a few false positives given that preprocessed sentences are used instead of original sentences.
However, its impact might be low since stemming is not performed that affects the individual words, and highly connected terms are chosen as search terms.   



\section{Related Work}\label{sec:related}
\textbf{Search Query Suggestion \& Reformulation:}
A number of studies in the literature attempt to support software developers in \emph{concept/feature/concern location} tasks using search query suggestion. They apply different lightweight heuristics \cite{kevic}, structural analyses \cite{twkraft,stacktrace} and query reformulation strategies \cite{gayg, refoqus,soniatool,shepherd}. They also perform different query quality analyses \cite{qquality,qeffect,specificity,soniaase} and data mining activities \cite{kevicdict,ccmapping,infer}. However, most of these approaches (1) expect a developer to provide the initial search query which they can improve upon, and (2) their main focus is improving a given query from the change request. 
Unfortunately, as existing studies \cite{kevic,vocaprob} suggest, preparing an initial search query is equally challenging, and those approaches do not provide much support in this regard.
In this study, we propose a novel technique--STRICT--that suggests a list of suitable terms as an \emph{initial search query} from a change request. 
\citet{kevic} consider a list of heuristics such as \emph{frequency, location, parts of speech} and \emph{notation} of the terms in the task description, and employ a prediction model for identifying search terms from a change request.
Our work is closely related to theirs, and we compare with it directly in our experiments.
In essence, our work is also aligned with query reformulation domain since it reformulates a baseline query by discarding the low quality search terms from the query.
According to \citet{refoqus}, Rocchio's method \cite{rocchio} is found to be one of the best query reformulation strategies, and thus is also closely related to ours. 
We also directly compare with this existing technique, and 
the detailed comparison can be found in Sections \ref{sec:comparison}.
\citeauthor{refoqus} and colleagues conduct several other studies on how to reformulate a given search query that apply machine learning \cite{refoqus} and query quality analyses \cite{qeffect,specificity}.
Given that a well-formed initial query requires less reformulation and preparing the initial query is already challenging \cite{kevic}, 
our technique can complement their techniques. 
\citet{ccmapping} suggest semantically similar query for a given query by mining comment-code mapping from a source code repository.
It could also possibly perform better if the initial query is prepared carefully which our technique does, rather than the query is chosen randomly.
\citet{twkraft} apply structural term weighting to feature location by emphasizing on source code tokens during query formulation.
However, as our finding suggests (RQ$_1$, Section \ref{sec:evaluation}), code tokens might not be always available, and thus queries based on them could be limited in performance.    
Other related studies apply ontology \cite{ontology}, query-based configurations \cite{trconfig,twkraft,stacktrace} and phrasal concepts \cite{ehill,shepherd} in concept location.

\textbf{Search Mechanisms:}
There also exist a number of studies on search mechanisms \cite{antoniol,irmarcus,fca,clustering,fusion,sitir} in the literature that apply the search queries to actually locate the concepts, features, concerns or bugs within the source code. These studies adopt static analyses, dynamic analyses or perform both analyses on the source code to identify the items of interest (\eg\ methods to be changed). \citet{fusion} combine information from three different processes--textual analysis, dynamic analysis and web mining--and apply \emph{PageRank} algorithm \cite{pagerank} like ours. However, they apply that algorithm in a different context--ranking methods within the project source--rather than search term identification from a change request.     
Other studies are not strongly related to our work--search term identification, and we did not compare with them directly in our experiments.

Thus, from a technical perspective, we adapt two information retrieval techniques--TextRank and POSRank--in the context of concept location, and identify a list of suitable search terms from a change request.
We exploit not only the co-occurrences but also the grammatical dependencies among the words from the request text carefully using a graph-based term weighting approach, and then capture the relative importance of those terms. 
Such idea was considered by no relevant existing studies, and the experimental findings also confirm the high potential of our idea.


\section{Conclusion \& Future Work\vspace{-.1cm}}
\label{sec:conclusion}
Studies suggest that preparing a search query from a change request is challenging for the developers.
In this paper, we propose a novel technique--STRICT--that automatically suggests suitable search terms from a change request.
It employs two term-weighting techniques--TextRank and POSRank-- from information-retrieval to capture a term's importance.
Experiments with 1,939 change requests from 8 subject systems show that our technique can identify better quality search terms than baseline for 52\%--62\% of the requests, and retrieves relevant results with 30\%--57\% Top-10 accuracy and about 30\% mean average precision, which are promising.
Comparison with two state-of-the-art techniques validates our empirical findings and also confirms the superiority of our technique.
In future, 
we plan to further extend our evaluation and implement the technique as an Eclipse plug-in tool.
Experimental data along with supporting materials are available elsewhere \cite{strict}.

\balance

\bibliographystyle{plainnat}
\setlength{\bibsep}{0pt plus 0.3ex}
\bibliography{sigproc}  

\begin{thebibliography}{48}
\providecommand{\natexlab}[1]{#1}
\providecommand{\url}[1]{\texttt{#1}}
\expandafter\ifx\csname urlstyle\endcsname\relax
  \providecommand{\doi}[1]{doi: #1}\else
  \providecommand{\doi}{doi: \begingroup \urlstyle{rm}\Url}\fi

\bibitem[str()]{strict}
{STRICT: Experimental Data}.
\newblock URL \url{http://homepage.usask.ca/~masud.rahman/strict}.

\bibitem[Antoniol et~al.(2002)Antoniol, Canfora, Casazza, De~Lucia, and
  Merlo]{antoniol}
G.~Antoniol, G.~Canfora, G.~Casazza, A.~De~Lucia, and E.~Merlo.
\newblock Recovering {T}raceability {L}inks between {C}ode and {D}ocumentation.
\newblock \emph{TSE}, 28\penalty0 (10):\penalty0 970--983, 2002.

\bibitem[Bachmann and Bernstein(2009)]{bugid}
A.~Bachmann and A.~Bernstein.
\newblock Software process data quality and characteristics: A historical view
  on open and closed source projects.
\newblock In \emph{Proc. IWPSE}, pages 119--128, 2009.

\bibitem[Bassett and Kraft(2013)]{twkraft}
B.~Bassett and N.~A. Kraft.
\newblock Structural information based term weighting in text retrieval for
  feature location.
\newblock In \emph{Proc. ICPC}, pages 133--141, 2013.

\bibitem[Bavota et~al.(2013)Bavota, De~Lucia, Oliveto, Panichella, Ricci, and
  Tortora]{bavota}
G.~Bavota, A.~De~Lucia, R.~Oliveto, A.~Panichella, F.~Ricci, and G.~Tortora.
\newblock The {R}ole of {A}rtefact {C}orpus in {LSI}-based {T}raceability
  {R}ecovery.
\newblock In \emph{Proc. TEFSE}, pages 83--89, 2013.

\bibitem[Blanco and Lioma(2012)]{blanco}
R.~Blanco and C.~Lioma.
\newblock Graph-based {T}erm {W}eighting for {I}nformation {R}etrieval.
\newblock \emph{Inf. Retr.}, 15\penalty0 (1):\penalty0 54--92, 2012.

\bibitem[Brin and Page(1998)]{pagerank}
S.~Brin and L.~Page.
\newblock The {A}natomy of a {L}arge-{S}cale {H}ypertextual {W}eb {S}earch
  {E}ngine.
\newblock \emph{Comput. Netw. ISDN Syst.}, 30\penalty0 (1-7):\penalty0
  107--117, 1998.

\bibitem[Capobianco et~al.(2013)Capobianco, Lucia, Oliveto, Panichella, and
  Panichella]{giovanni}
G.~Capobianco, A.~D. Lucia, R.~Oliveto, A.~Panichella, and S.~Panichella.
\newblock {Improving IR-based Traceability Recovery via Noun-Based Indexing of
  Software Artifacts}.
\newblock \emph{Journal of Software: Evolution and Process}, 25\penalty0
  (7):\penalty0 743--762, 2013.

\bibitem[Cordeiro et~al.(2012)Cordeiro, Antunes, and Gomes]{context}
J.~Cordeiro, B.~Antunes, and P.~Gomes.
\newblock {C}ontext-based {R}ecommendation to {S}upport {P}roblem {S}olving in
  {S}oftware {D}evelopment.
\newblock In \emph{Proc. RSSE}, pages 85 --89, 2012.

\bibitem[Dagenais and Robillard(2010)]{ossdoc}
B.~Dagenais and M.~P. Robillard.
\newblock {Creating and Evolving Developer Documentation: Understanding the
  Decisions of Open Source Contributors}.
\newblock In \emph{Proc. FSE}, pages 127--136, 2010.

\bibitem[Dit et~al.(2011)Dit, Guerrouj, Poshyvanyk, and Antoniol]{splitting}
B.~Dit, L.~Guerrouj, D.~Poshyvanyk, and G.~Antoniol.
\newblock Can better identifier splitting techniques help feature location?
\newblock In \emph{Proc. ICPC}, pages 11--20, 2011.

\bibitem[Furnas et~al.(1987)Furnas, Landauer, Gomez, and Dumais]{vocaprob}
G.~W. Furnas, T.~K. Landauer, L.~M. Gomez, and S.~T. Dumais.
\newblock {The Vocabulary Problem in Human-system Communication}.
\newblock \emph{Commun. ACM}, 30\penalty0 (11):\penalty0 964--971, 1987.

\bibitem[Gay et~al.(2009)Gay, Haiduc, Marcus, and Menzies]{gayg}
G.~Gay, S.~Haiduc, A.~Marcus, and T.~Menzies.
\newblock On the {U}se of {R}elevance {F}eedback in {IR}-based {C}oncept
  {L}ocation.
\newblock In \emph{Proc. ICSM}, pages 351--360, 2009.

\bibitem[Haiduc(2011)]{qquality}
S.~Haiduc.
\newblock Automatically {D}etecting the {Q}uality of the {Q}uery and its
  {I}mplications in {IR}-based {C}oncept {L}ocation.
\newblock In \emph{Proc. ASE}, pages 637--640, 2011.

\bibitem[Haiduc and Marcus(2011)]{qeffect}
S.~Haiduc and A.~Marcus.
\newblock {On the Effect of the Query in IR-based Concept Location}.
\newblock In \emph{Proc. ICPC}, pages 234--237, 2011.

\bibitem[Haiduc et~al.(2012{\natexlab{a}})Haiduc, Bavota, Oliveto, Lucia, and
  Marcus]{soniaase}
S.~Haiduc, G.~Bavota, R.~Oliveto, A.~De Lucia, and A.~Marcus.
\newblock {Automatic query performance assessment during the retrieval of
  software artifacts}.
\newblock In \emph{Proc. ASE}, pages 90--99, 2012{\natexlab{a}}.

\bibitem[Haiduc et~al.(2012{\natexlab{b}})Haiduc, Bavota, Oliveto, Marcus, and
  Lucia]{specificity}
S.~Haiduc, G.~Bavota, R.~Oliveto, A.~Marcus, and A.~De Lucia.
\newblock {Evaluating the Specificity of Text Retrieval Queries to Support
  Software Engineering Tasks}.
\newblock In \emph{Proc. ICSE}, pages 1273--1276, 2012{\natexlab{b}}.

\bibitem[Haiduc et~al.(2013{\natexlab{a}})Haiduc, Bavota, Marcus, Oliveto,
  De~Lucia, and Menzies]{refoqus}
S.~Haiduc, G.~Bavota, A.~Marcus, R.~Oliveto, A.~De~Lucia, and T.~Menzies.
\newblock Automatic {Q}uery {R}eformulations for {T}ext {R}etrieval in
  {S}oftware {E}ngineering.
\newblock In \emph{Proc. ICSE}, pages 842--851, 2013{\natexlab{a}}.

\bibitem[Haiduc et~al.(2013{\natexlab{b}})Haiduc, De~Rosa, Bavota, Oliveto,
  De~Lucia, and Marcus]{soniatool}
S.~Haiduc, G.~De~Rosa, G.~Bavota, R.~Oliveto, A.~De~Lucia, and A.~Marcus.
\newblock {Query Quality Prediction and Reformulation for Source Code Search:
  The Refoqus Tool}.
\newblock In \emph{Proc. ICSE}, pages 1307--1310, 2013{\natexlab{b}}.

\bibitem[Hassan et~al.(2007)Hassan, Mihalcea, and Banea]{sameer}
S.~Hassan, R.~Mihalcea, and C.~Banea.
\newblock {Random-Walk Term Weighting for Improved Text Classification}.
\newblock In \emph{Proc. ICSC}, pages 242--249, 2007.

\bibitem[Hill et~al.(2011)Hill, Pollock, and Vijay-Shanker]{ehill}
E.~Hill, L.~Pollock, and K.~Vijay-Shanker.
\newblock {Improving Source Code Search with Natural Language Phrasal
  Representations of Method Signatures}.
\newblock In \emph{Proc. ASE}, pages 524--527, 2011.

\bibitem[Howard et~al.(2013)Howard, Gupta, Pollock, and
  Vijay-Shanker]{ccmapping}
M.~J. Howard, S.~Gupta, L.~Pollock, and K.~Vijay-Shanker.
\newblock Automatically {M}ining {S}oftware-based, {S}emantically-{S}imilar
  {W}ords from {C}omment-{C}ode {M}appings.
\newblock In \emph{Proc. MSR}, pages 377--386, 2013.

\bibitem[Jespersen(1929)]{jespersen}
Otto Jespersen.
\newblock {The Philosophy of Grammar}.
\newblock 1929.

\bibitem[Kevic and Fritz(2014{\natexlab{a}})]{kevic}
K.~Kevic and T.~Fritz.
\newblock Automatic {S}earch {T}erm {I}dentification for {C}hange {T}asks.
\newblock In \emph{Proc. ICSE}, pages 468--471, 2014{\natexlab{a}}.

\bibitem[Kevic and Fritz(2014{\natexlab{b}})]{kevicdict}
K.~Kevic and T.~Fritz.
\newblock A {D}ictionary to {T}ranslate {C}hange {T}asks to {S}ource {C}ode.
\newblock In \emph{Proc. MSR}, pages 320--323, 2014{\natexlab{b}}.

\bibitem[Ko et~al.(2006)Ko, Myers, Coblenz, and Aung]{topten}
Andrew~J. Ko, Brad~A. Myers, Michael~J. Coblenz, and Htet~Htet Aung.
\newblock {An Exploratory Study of How Developers Seek, Relate, and Collect
  Relevant Information During Software Maintenance Tasks}.
\newblock \emph{TSE}, 32\penalty0 (12):\penalty0 971--987, 2006.

\bibitem[Liu et~al.(2007)Liu, Marcus, Poshyvanyk, and Rajlich]{sitir}
D.~Liu, A.~Marcus, D.~Poshyvanyk, and V.~Rajlich.
\newblock Feature location via information retrieval based filtering of a
  single scenario execution trace.
\newblock In \emph{Proc. ASE}, pages 234--243, 2007.

\bibitem[Marcus and Haiduc(2013)]{textret}
A.~Marcus and S.~Haiduc.
\newblock Text {R}etrieval {A}pproaches for {C}oncept {L}ocation in {S}ource
  {C}ode.
\newblock In \emph{Software Engineering}, volume 7171, pages 126--158. 2013.

\bibitem[Marcus et~al.(2004)Marcus, Sergeyev, Rajlich, and Maletic]{irmarcus}
A.~Marcus, A.~Sergeyev, V.~Rajlich, and J.I. Maletic.
\newblock An {I}nformation {R}etrieval {A}pproach to {C}oncept {L}ocation in
  {S}ource {C}ode.
\newblock In \emph{Proc. WCRE}, pages 214--223, 2004.

\bibitem[Mihalcea and Tarau(2004)]{rada}
R.~Mihalcea and P.~Tarau.
\newblock Textrank: {B}ringing {O}rder into {T}exts.
\newblock In \emph{Proc. EMNLP}, pages 404--411, 2004.

\bibitem[Moreno et~al.(2014)Moreno, Treadway, Marcus, and Shen]{stacktrace}
L.~Moreno, J.~J. Treadway, A.~Marcus, and W.~Shen.
\newblock On the use of stack traces to improve text retrieval-based bug
  localization.
\newblock In \emph{Proc. ICSME}, pages 151--160, 2014.

\bibitem[Moreno et~al.(2015)Moreno, Bavota, Haiduc, Di~Penta, Oliveto, Russo,
  and Marcus]{trconfig}
L.~Moreno, G.~Bavota, S.~Haiduc, M.~Di~Penta, R.~Oliveto, B.~Russo, and
  A.~Marcus.
\newblock Query-based configuration of text retrieval solutions for software
  engineering tasks.
\newblock In \emph{Proc. ESEC/FSE}, pages 567--578, 2015.

\bibitem[Parnin and Treude(2011)]{measure}
C.~Parnin and C.~Treude.
\newblock {Measuring API Documentation on the Web}.
\newblock In \emph{Proc. Web2SE}, pages 25--30, 2011.

\bibitem[Poshyvanyk and Marcus(2007)]{fca}
D.~Poshyvanyk and A.~Marcus.
\newblock Combining {F}ormal {C}oncept {A}nalysis with {I}nformation
  {R}etrieval for {C}oncept {L}ocation in {S}ource {C}ode.
\newblock In \emph{Proc. ICPC}, pages 37--48, 2007.

\bibitem[Rahman and Roy(2015)]{saner2015masud}
M.~M. Rahman and C.~K. Roy.
\newblock {TextRank Based Search Term Identification for Software Change
  Tasks}.
\newblock In \emph{Proc. SANER}, pages 540--544, 2015.

\bibitem[Rahman et~al.(2014)Rahman, Yeasmin, and Roy]{surfclipse}
M.~M. Rahman, S.~Yeasmin, and C.~K. Roy.
\newblock Towards a {C}ontext-{A}ware {IDE}-{B}ased {M}eta {S}earch {E}ngine
  for {R}ecommendation about {P}rogramming {E}rrors and {E}xceptions.
\newblock In \emph{Proc. CSMR-WCRE}, pages 194--203, 2014.

\bibitem[Rahman et~al.(2016)Rahman, Roy, and Lo]{saner2016masud}
M.~M. Rahman, C.~K. Roy, and D.~Lo.
\newblock {RACK}: {A}utomatic {API} {R}ecommendation using {C}rowdsourced
  {K}nowledge.
\newblock In \emph{Proc. SANER}, pages 349--359, 2016.

\bibitem[Revelle et~al.(2010)Revelle, Dit, and Poshyvanyk]{fusion}
M.~Revelle, B.~Dit, and D.~Poshyvanyk.
\newblock Using data fusion and web mining to support feature location in
  software.
\newblock In \emph{Proc. ICPC}, pages 14--23, 2010.

\bibitem[Rocchio()]{rocchio}
J.J. Rocchio.
\newblock \emph{The SMART Retrieval System---Experiments in Automatic Document
  Processing}.
\newblock Prentice-Hall, Inc.

\bibitem[Saha et~al.(2013)Saha, Lease, Khurshid, and Perry]{saha}
R.~K. Saha, M.~Lease, S.~Khurshid, and D.~E. Perry.
\newblock Improving bug localization using structured information retrieval.
\newblock In \emph{Proc. ASE}, pages 345--355, 2013.

\bibitem[Scanniello and Marcus(2011)]{clustering}
G.~Scanniello and A.~Marcus.
\newblock {Clustering Support for Static Concept Location in Source Code}.
\newblock In \emph{Proc. ICPC}, pages 1--10, 2011.

\bibitem[Shepherd et~al.(2007)Shepherd, Fry, Hill, Pollock, and
  Vijay-Shanker]{shepherd}
D.~Shepherd, Z.~P. Fry, E.~Hill, L.~Pollock, and K.~Vijay-Shanker.
\newblock Using {N}atural {L}anguage {P}rogram {A}nalysis to {L}ocate and
  {U}nderstand {A}ction-{O}riented {C}oncerns.
\newblock In \emph{Proc. ASOD}, pages 212--224, 2007.

\bibitem[Thongtanunam et~al.(2015)Thongtanunam, Kula, Yoshida, Iida, and
  Matsumoto]{pick}
P.~Thongtanunam, R.~G. Kula, N.~Yoshida, H.~Iida, and K.~Matsumoto.
\newblock Who {S}hould {R}eview my {C}ode ?
\newblock In \emph{Proc. SANER}, pages 141--150, 2015.

\bibitem[Toutanova et~al.(2003)Toutanova, Klein, Manning, and
  Singer]{postagger}
K.~Toutanova, D.~Klein, C.D. Manning, and Y.~Singer.
\newblock {Feature-Rich Part-of-Speech Tagging with a Cyclic Dependency
  Network}.
\newblock In \emph{Proc. HLT-NAACL}, pages 252--259, 2003.

\bibitem[Wilson(2010)]{ontology}
L.A. Wilson.
\newblock Using {O}ntology {F}ragments in {C}oncept {L}ocation.
\newblock In \emph{Proc. ICSM}, pages 1--2, 2010.

\bibitem[Yang and Tan(2012)]{infer}
J.~Yang and L.~Tan.
\newblock {Inferring Semantically Related Words from Software Context}.
\newblock In \emph{Proc. MSR}, pages 161--170, 2012.

\bibitem[Yuan et~al.(2014)Yuan, Lo, and Lawall]{wordsim}
T.~Yuan, D.~Lo, and J.~Lawall.
\newblock Automated {C}onstruction of a {S}oftware-specific {W}ord {S}imilarity
  {D}atabase.
\newblock In \emph{Proc. CSMR-WCRE}, pages 44--53, 2014.

\bibitem[Zhou et~al.(2012)Zhou, Zhang, and Lo]{buglocator}
J.~Zhou, H.~Zhang, and D.~Lo.
\newblock Where should the bugs be fixed? - more accurate information
  retrieval-based bug localization based on bug reports.
\newblock In \emph{Proc. ICSE}, 2012.

\end{thebibliography}
%
%

\end{document}